\documentclass[conference,10pt]{IEEEtran}
\usepackage{graphicx}
\usepackage{amsfonts,color}
\usepackage{cite}
\usepackage{epsfig}
\usepackage{latexsym}
\usepackage{stmaryrd}
\usepackage[T1]{fontenc}
\usepackage{epstopdf}
\usepackage{amsthm, amssymb}
\usepackage{verbatim}
\usepackage{amsmath}
\usepackage{algorithmic}
\usepackage[algo2e]{algorithm2e}
\usepackage{algorithm}
\usepackage{subcaption}
\usepackage{soul}

\newtheorem{theorem}{Theorem}

\newtheorem{remark}{Remark}
\newtheorem{definition}{Definition}
\definecolor{orange}{RGB}{255,107,0}

\IEEEoverridecommandlockouts

\begin{document}
		\title{ \fontsize{19}{27}\selectfont Vandermonde Constrained Tensor Decomposition for Hybrid Beamforming in Multi-Carrier MIMO Systems}
		\author{\IEEEauthorblockN{Mohamed Salah Ibrahim$^\dagger$, Akshay Malhotra$^{\ddagger}$, Mihaela Beluri$^*$, Arnab Roy$^\dagger$, and Shahab Hamidi-Rad$^{\ddagger}$}
		\IEEEauthorblockA{$^\dagger$InterDigital Communications Inc., Conshohocken, PA, USA, 	$^{\ddagger}$InterDigital Communications Inc., Los Altos, CA, USA\\ 	$^*$InterDigital Communications Inc., New York, NY, USA \\
			                   Email: \{\tt mohamedsalah.ibrahim,akshay.malhotra,mihaela.beluri,\\arnab.roy,shahab.hamidi-rad@interdigital.com\}
		                 }
	           }

\maketitle
\begin{abstract}
Hybrid beamforming has evolved as a promising technology that offers the balance between system performance and design complexity in mmWave MIMO systems. Existing hybrid beamforming methods either impose unit-modulus constraints or a codebook constraint on the analog precoders/combiners, which in turn results in a performance-overhead tradeoff. This paper puts forth a tensor framework to handle the wideband hybrid beamforming problem, with Vandermonde constraints on the analog precoders/combiners. The proposed method strikes the balance between performance, overhead and complexity. Numerical results on a 3GPP link-level test bench reveal the efficacy of the proposed approach relative to the codebook-based method while attaining the same feedback overhead. Moreover, the proposed method is shown to achieve comparable performance to the unit-modulus approaches, with substantial reductions in overhead.
\end{abstract}

\section{introduction}\label{Intro}
Millimeter wave (mmWave) has emerged as a powerful technology, that can handle the unprecedented demands on wireless connectivity, through offering large available bandwidth~\cite{boccardi2014five}. However, the high propagation loss inherent to mmWave bands, if not mitigated, can severely impact the system performance. Large antenna arrays which achieve high beamforming gains are used to compensate the propagation loss~\cite{rangan2014millimeter}. 


Large scale antenna systems implementation, on the other hand, incurs several practical challenges including the high energy consumption and cost of radio frequency (RF) chains, as each antenna element requires a dedicated RF chain. Such hurdles limit the possibility of employing a fully digital beamforming design. As an {\em efficient} surrogate, hybrid (analog/digital) beamforming has been introduced in \cite{el2014spatially,alkhateeb2014channel} as means of attaining favorable complexity-performance tradeoff in mmWave multicarrier massive MIMO systems. Hybrid beamforming relies on using a small number of RF chains to design high-dimensional analog precoders (implemented with only phase shifters) together with a low-dimensional (digital) baseband precoder. The combination of analog and digital precoders has the potential to approach the performance of a purely digital solution while providing substantial savings in energy consumption and design complexity.  

Although maximizing the system spectral efficiency in the case of digital beamforming design admits a simple algebraic solution via singular value decomposition (SVD) \cite{goldsmith2003capacity}, hybrid beamforming yields a highly non-convex problem that requires joint optimization of the hybrid precoders and combiners \cite{el2014spatially}. A more tractable formulation is to transform the hybrid beamforming design to a matrix factorization problem. In particular, the optimal SVD-based  digital solution is first derived to maximize the spectral efficiency. Then, the hybrid beamforming is posed towards factorizing the fully digital precoder (combiner) as the hybrid precoding (combining) components. The factorization is usually solved either under unit modulus constraints \cite{el2014spatially, yu2016alternating} or with codebook constraints \cite{alkhateeb2016frequency} on the analog precoder (combiner), to ensure that the analog precoder can be modeled using phase shifters. While considering the unit modulus constraints, in general, result in a much better solution compared to the codebook constraint \cite{yu2016alternating}, the resulting communication overhead of the latter is considerably lower \cite{alkhateeb2016frequency}, rendering it more appropriate for limited feedback systems \cite{love2008overview}. Further, compared to the codebook constraints approach, the feedback overhead for unit magnitude constraints scales linearly with the number of Tx/Rx antennas, thereby precluding its use in massive MIMO systems. 

This begs the question whether it is possible to achieve a comparable performance to the unit-modulus based methods while yielding the feedback  associated with the codebook-based approaches? This is the central question that this paper seeks to address. We answer the stated question in the affirmative by modeling the wideband hybrid beamforming as a low rank tensor decomposition problem with Vandermonde constraints on the analog precoders/combiners. Invoking the so-called parallel factor (PARAFAC) analysis, to decompose the resulting tensor, we show that PARAFAC yields high-quality hybrid precoders/combiners, with identifiability guarantees on the resulting factors. This paper adds to the broad variety of tensors applications in wireless communications \cite{de2007parafac,sidiropoulos2000blind,ibrahim2020downlink}. Different from all prior hybrid beamforming works that adopt the spectral efficiency formula for performance evaluation, this paper evaluates the practical impact of the proposed method by integrating hybrid beamforming to an end-to-end communication scenario with time-varying channels. Numerical results demonstrate that the end-to-end performance of the proposed approach considerably outperforms the codebook based method while achieving comparable performance to the unit modulus based approaches. Further, the proposed method yields significantly lower communication overhead compared to unit modulus approaches.


\vspace{-0.2cm}
\section{System Model}\label{Sec:SysModel}
\begin{figure}
	\centering
	\includegraphics[width=.5\textwidth]{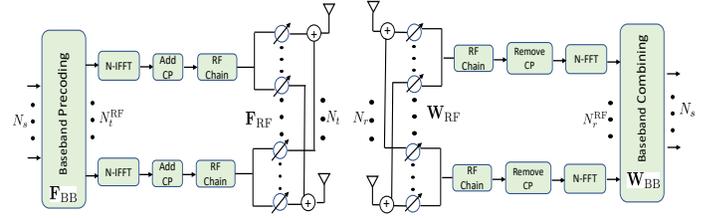}
	\caption{Block diagram of OFDM-MIMO system with a BS and a UE employing hybrid precoding and combining.}
	\label{Fig:SysModel}
	\end{figure}

Consider a downlink transmission in a multi-carrier MIMO system comprising a base station (BS) and a single user equipment (UE). The BS is equipped with $N_t$ transmit antennas and $N_t^\text{RF}$ transmit radio frequency (RF) chains while the UE is equipped with $N_r$ receive antennas and $N_r^\text{RF}$ receive RF chains. The BS aims at communicating $N_s$ data streams to the UE over $K$ subcarriers, where $N_s \leq N_t^\text{RF} \leq N_t$ and $N_s \leq N_r^\text{RF} \leq N_r$ \cite{el2014spatially}. The BS first employs a digital baseband precoding matrix ${\bf F}_\text{BB}[k] \in \mathbb{C}^{N_t^\text{RF} \times N_s}$ on the transmitted symbols ${\bf s}[k] \in \mathbb{C}^{N_s}$, $\forall k \in [K] := \{0,\cdots,K-1\}$, as shown in Fig. \ref{Fig:SysModel}. Then, the data symbols are transformed to the time domain using N-point inverse fast Fourier transform (IFFT). After a cyclic prefix (CP) is added to the time-domain signal, the BS applies an analog precoder ${\bf F}_\text{RF} \in \mathbb{C}^{N_t \times N_t^\text{RF}}$ (implemented using analog phase shifters), i.e., $|{\bf F}_\text{RF}(i,j)| =1,~\forall i=0,\cdots,N_t-1~\text{and}~j=0,\cdots,N_t^\text{RF}-1$. Notice that same ${\bf F}_\text{RF}$ is applied across all subcarriers, i.e., ${\bf F}_\text{RF}$ is frequency independent. Towards this end, the transmitted complex signal from the BS can be expressed as,
\begin{align}\label{Eq:TxSig}
    {\bf x}[k] = {\bf F}_\text{RF}{\bf F}_\text{BB}[k]{\bf s}[k],~\forall k \in [K].
\end{align}
It is assumed that i) $\mathbb{ E}[{\bf s}[k]{\bf s}^H[k]] = \frac{\alpha}{KN_s}{\bf I}_{N_s}$, and ii) the total power budget constraint ${\alpha}$ is satisfied by enforcing the constraint $\| {\bf F}_\text{RF}{\bf F}_\text{BB}[k] \|^2_F = N_s~\forall k \in [K]$.

At the receiver, the UE first employs an analog combiner ${\bf W}_\text{RF} \in \mathbb{C}^{N_r \times N_r^\text{RF}} $ followed by a digital baseband combiner ${\bf W}_\text{BB} \in \mathbb{C}^{N_r^\text{RF} \times N_s}$ after CP removal and frequency transformation using N-point FFT. Similar to the unit modulus constraint on the entries of ${\bf F}_\text{RF}$, it assumed that the $(i,j)$-th entry of ${\bf W}_\text{RF}$ has a unit modulus, i.e., $|{\bf W}_\text{RF}(i,j)| = 1,~\forall i=0,\cdots,N_r^\text{RF}-1~\text{and}~j=0,\cdots,N_r-1$. Thus, the $N_s$-dimensional complex baseband signal at the UE at the $k$-th subcarrier is given by,
\begin{align}\label{Eq:RecSig}
    {\bf y}[k] = {\bf W}^H_\text{BB}[k]{\bf W}^H_\text{RF}{\bf H}[k]{\bf x}[k] +  {\bf W}^H_\text{BB}[k]{\bf W}^H_\text{RF}{\bf v}[k].
\end{align}
where ${\bf H}[k] \in \mathbb{C}^{N_r \times N_t}$ represents the downlink channel at the $k$-th subcarrier, and ${\bf v}[k] \in \mathbb{C}^{N_r}$ is the additive white Guassian noise vector associated with the $k$-th subcarrier,$~\forall k \in [K]$. It is assumed that the entries of ${\bf v}[k]$ are independent and identically distributed (i.i.d) random variables with zero mean and variance $\sigma^2$, i.e., ${\bf v}[k] \sim  \mathcal{N}(0,\sigma^2{\bf I}_{N_r}),~\forall i = 0,\cdots,N_r-1$. Throughout this work, we assume that the channel matrices across the subcarriers $\{{\bf H}[k]\}_{k=1}^{K}$ are perfectly known at the UE.
\begin{remark}
    It is worth pointing out that, in practical wireless systems, there is one representative channel matrix for each group of subcarriers or resource blocks, referred to as subband size, and hence, there is one baseband precoder/combiner for each subband as opposed to each subcarrier. The reason behind that is primarily to reduce the overhead associated with the channel and/or precoding-related feedback. In the hybrid beamforming context, this will reduce the overhead associated with the baseband precoders/combiners, and will also reduce the complexity as obviously smaller number of baseband precoders and combiners need to be computed. This fact will be utilized later in the simulations (Section \ref{Sec:Simu}) and also in the overhead computations associated with the proposed approach and the existing hybrid beamforming methods.   
\end{remark}
	\vspace{-0.3cm}
\section{Problem Definition}\label{ProbDef}
The wideband hybrid beamforming problem seeks to find the set of hybrid precoders and combiners $({\bf F}_\text{RF},\{{\bf F}_\text{BB}[k]\}_{k=1}^K,{\bf W}_\text{RF},\{{\bf W}_\text{BB}[k]\}_{k=1}^{K})$ that can maximize the spectral efficiency. Assuming the transmitted symbols follow a Gaussian distribution, the achievable spectral efficiency associated with the $k$-th subcarrier can be expressed as~\cite{goldsmith2003capacity}
\begin{align}\label{Eq:ShannonFormula}
          \text{R}[k] &= \log_2(\text{det}({\bf I}_{N_s} + \frac{\alpha}{N_s}{\bf \Gamma}^{-1}[k]{\bf W}^H_\text{BB}[k]{\bf W}^H_\text{RF}{\bf H}[k]{\bf F}_\text{RF}{\bf F}_\text{BB}[k] \nonumber \\  &\times {\bf F}^H_\text{BB}[k]{\bf F}^H_\text{RF}{\bf H}^H[k]{\bf W}_\text{RF}{\bf W}_\text{BB}[k]))
\end{align}
where ${\bf \Gamma}[k] := \sigma^2 {\bf W}^H_\text{BB}[k]{\bf W}^H_\text{RF}{\bf W}_\text{RF}{\bf W}_\text{BB}[k]$ represents the covariance matrix of the post-processing noise term in~\eqref{Eq:RecSig}. The goal is then to design the hybrid precoders $({\bf F}_\text{RF},\{{\bf F}_\text{BB}[k]\}_{k=1}^K,{\bf W}_\text{RF},\{{\bf W}_\text{BB}[k]\}_{k=1}^{K})$ that aim at maximizing the overall spectral efficiency while satisfying the imposed constraints on the analog and digital precoders/combiners. Maximizing the spectral efficiency, though, yields a highly non tractable optimization problem that requires the hybrid precoders and combiners to be jointly optimized. 

Instead of maximizing the spectral efficiency, one can decouple the precoders and combiners design, and formulate the hybrid beamforming problem as two separate low-rank matrix factorization problems~\cite{el2014spatially,yu2016alternating,el2012low}. The precoder problem aims at factorizing the optimal digital precoder ${\bf F}_\text{opt}[k] \in \mathbb{C}^{N_t \times N_s}$ to ${\bf F}_\text{RF}{\bf F}_\text{BB}[k]$, where the columns of ${\bf F}_\text{opt}[k]$ are the $N_s$ dominant right singular vectors of ${\bf H}[k]$, $\forall k \in [K]$. On the other hand, the combiner problem seeks to factorize  ${\bf W}_\text{opt}[k] \in \mathbb{C}^{N_r \times N_s}$ to ${\bf W}_\text{RF}{\bf W}_\text{BB}[k]$, where ${\bf W}_\text{opt}[k]$ is the WMMSE solution, i.e., ${\bf W}_\text{opt}[k] = (\overline{\bf H}^H[k]\overline{\bf H}[k] + \sigma^2 {\bf I}_{N_s})^{-1}\overline{\bf H}^H$ and $\overline{\bf H}[k] = {\bf H}[k]{\bf F}_\text{opt}[k] $ . Interestingly, it has been shown in \cite{el2012low} that solving the factorization problems implicitly leads to maximizing the system spectral efficiency. Since both problems exhibit similar mathematical formulation, except that the precoder problem has an additional sum power constraint, we will focus on the precoder factorization problem. However, the proposed method may be easily applied to solve the combiner problem. From an optimization perspective, given the fully digital SVD-based precoder ${\bf F}_\text{opt}[k]$, the hybrid beamforming problem can be posed as~\cite{el2014spatially,yu2016alternating,el2012low}
\begin{subequations}\label{OP:WBHBF}
	\begin{align}
   & \min_{{\bf F}_\text{RF},\{{\bf F}_\text{BB}[k]\}_{k=1}^K} \sum\limits_{k=1}^{K}\| {\bf F}_\text{opt}[k] - {\bf F}_\text{RF}{\bf F}_\text{BB}[k] \|_F^2\\ 
    &\textrm{s.t.} ~~~~~~~~~~~~~~ {\bf F}_\text{RF} \in \mathcal{F}\\
        & ~~~~~~~~~~~~~~~~\| {\bf F}_\text{RF}{\bf F}_\text{BB}[k] \|^2_F = N_s,~\forall k \in [K]
	\end{align}
\end{subequations}
where $\mathcal{F}$ is the feasible set of the analog precoders. In the wideband hybrid beamforming literature, the feasible set either includes unit modulus constraints on the entries of ${\bf F}_\text{RF}$~\cite{yu2016alternating} (denoted as $\mathcal{F}_U$), or code-book based selection of the columns of ${\bf F}_\text{RF}$~\cite{alkhateeb2016frequency} (denoted as $\mathcal{F}_C$). 
The two feasible sets yield an interesting overhead-performance trade-off. While considering the feasible set $\mathcal{F}_C$ results in much lower overhead relative to the set $\mathcal{F}_U$, the solution associated with $\mathcal{F}_U$ performs much better than that of $\mathcal{F}_C$. The intuition is that $\mathcal{F}_U$ provides a much wider search space compared to $\mathcal{F}_C$, i.e., $\mathcal{F}_C \subset \mathcal{F}_U$, and hence, better performance is expected.

In this paper, we will introduce a new feasible set (denoted as $\mathcal{F}_V$) to the wideband hybrid beamforming problem in \eqref{OP:WBHBF}  by enforcing a Vandermonde structure on the columns of ${\bf F}_\text{RF}$, i.e., $\mathcal{F}_V := \{{\bf x} \in \mathbb{C}^{N_t}~ |~ {\bf x} = [1,e^{j\phi},\cdots,e^{j(N_t -1)\phi}]^T\}$, and $\phi \in [-\pi,\pi]$. Towards this end, the problem that this paper seeks to solve is the following low-rank matrix optimization problem,
\begin{subequations}\label{OP:WBHBFVan}
	\begin{align}
    &\text{Find} \quad  {\bf F}_{\text{RF}}(\phi_0,\cdots,\phi_{N_\text{RF}-1}),~ \{{\bf F}_{\text{BB}}[k]\}_{k=1}^{K}  \\
    &\textrm{s.t.} \quad~~ {\bf F}_{\text{opt}}[k] \approx {\bf F}_{\text{RF}}{\bf F}_{\text{BB}}[k],~{\bf F}_\text{RF} \in \mathcal{F}_V,~~\forall~ k \in [K].
	\end{align}
\end{subequations}
Notice that the sum power constraint in (\ref{OP:WBHBF}c) is temporarily omitted as it has been shown that such a constraint can be satisfied via a simple normalization step to the resulting baseband precoders \cite{el2014spatially}. To our best knowledge, the formulation in \eqref{OP:WBHBFVan} has not been considered before in the hybrid beamforming literature. Such a formulation strikes the balance between the obtained solution quality and the resulting overhead. In particular, the resulting solution achieves the same overhead associated with the $\mathcal{F}_C$ set while achieving comparable performance to the solutions associated with the $\mathcal{F}_U$ set. In the subsequent section, we will show that \eqref{OP:WBHBFVan} can be reformulated as a tensor factorization problem where efficient tensor decomposition methods can be applied. 

\vspace{-0.2cm}
\section{PARAFAC Decomposition}\label{Sec:TenandPropMethod}
Before reformulating \eqref{OP:WBHBFVan} as a tensor factorization problem and to facilitate our discussion, we briefly review some key concepts that will be used in the proposed tensor approach.
\vspace{-0.3cm}
\subsection{Tensor Preliminaries}\label{SubSec:TenPre}
A third order tensor $\mathcal{X} \in \mathbb{C}^{I \times J \times P}$ is a three way array whose elements are indexed by three indices $(i,j,p)$. The so-called  Parallel Factor decomposition (PARAFAC), a.k.a Canonical Polyadic Decomposition (CPD), is one  powerful tensor decomposition method. A tensor $\mathcal{X}$ admits a PARAFAC decomposition if it can be written as the sum of vector outer products  \cite{sidiropoulos2000uniqueness},
\begin{align}\label{TensorDecomOuter}
    \mathcal{X} = \sum\limits_{f = 1}^{F} {\bf a}_f \circ {\bf b}_f \circ {\bf c}_f.
\end{align}
where $\circ$ denotes the vector outer product, and $F$ is a positive integer that we refer to as the tensor rank or CPD rank (the smallest value such that \eqref{TensorDecomOuter} holds). The terms ${\bf a}_f \in \mathbb{C}^I$, ${\bf b}_f \in \mathbb{C}^J$ and ${\bf c}_f \in \mathbb{C}^P$ are the $f$-th columns of the so-called low-rank factors ${\bf A} \in \mathbb{C}^{I \times F}$, ${\bf B} \in \mathbb{C}^{J \times F}$, and ${\bf C} \in \mathbb{C}^{P \times F}$, respectively, of the tensor $\mathcal{X}$.

Different from the tensor format in \eqref{TensorDecomOuter}, PARAFAC can also be written in slab format. Let ${\bf X}_p := \mathcal{X}(:,:,p)$ represent the $p$-th frontal slab of $\mathcal{X},~\forall p \in [P] := \{0,\cdots,P-1\} $.\footnote{Note that we used the MATLAB notation $\mathcal{X}(:,:,p)$ to read the frontal slab of a three-way tensor.} The PARAFAC decomposition of $\mathcal{X}$ in the slab-format is given by
\begin{align}\label{TensorDecomSlab}
       \mathcal{X}(:,:,p) = {\bf A}{\bf D}_p({\bf C}){\bf B}^T,~~ \forall p \in [P].
\end{align}
where ${\bf D}_p({\bf C}) := \text{Diag}({\bf C}(p,:)) \in \mathbb{C}^{F \times F}$ with the elements on the diagonal be the $p$-th row of ${\bf C}$. Throughout this paper, we will use the notation $\mathcal{X} := \left\llbracket{\bf A},{\bf B},{\bf C}\right\rrbracket$ to denote \eqref{TensorDecomSlab}.

\subsection{Identifiability}
One distinctive property of tensors is that the PARAFAC model is essentially unique under mild conditions even if $F$ is greater than $\max(I,J,P)$. The definition of essential uniqueness is presented as follows.
\begin{definition}
The PARAFAC decomposition of a tensor $\mathcal{X}$ is said to be  {\em essentially unique}, $\mathcal{X}  := \left\llbracket{\bf A},{\bf B},{\bf C}\right\rrbracket$, if ${\bf A}, {\bf B}$ and ${\bf C}$ are identifiable up to scaling and permutation. This means that if  $\mathcal{X} := \left\llbracket \overline{\bf A},\overline{\bf B},\overline{\bf C} \right\rrbracket$, for some $\overline{\bf A} \in \mathbb{C}^{I \times F}$, $\overline{\bf B} \in \mathbb{C}^{J \times F}$, and $\overline{\bf C} \in \mathbb{C}^{P \times F}$, then there exists a permutation matrix ${\bf \Pi} \in \mathbb{R}^{F \times F}$ and diagonal scaling matrices $\{{\bf \Lambda}_i\}_{i=1}^{3}$ such that,
\begin{align}\label{TensPermutaionandScaling}
    {\bf A} = \overline{\bf A}{\bf \Pi}{\bf \Lambda}_1,    {\bf B} = \overline{\bf B}{\bf \Pi}{\bf \Lambda}_2,{\bf C} = \overline{\bf C}{\bf \Pi}{\bf \Lambda}_3, {\bf \Lambda}_1{\bf \Lambda}_2{\bf \Lambda}_3 = {\bf I}.
\end{align}
\end{definition}

If there is no structure imposed on the low rank factors, then a generic identifiability condition on PARAFAC uniqueness is given in \cite{chiantini2012generic}. If, however, one or more of the low rank factor matrices have a Vandermonde structure, then more relaxed uniqueness conditions based on the Kruskal rank can be found in~\cite{sidiropoulos2000uniqueness,de2006link,sidiropoulos2000blind}. The latest and the most relevant identifiability results to the problem considered herein is given as follows. 

\begin{theorem}
\cite{sidiropoulos2000parallel} Consider the data model in \eqref{TensorDecomSlab} and assume that the factors ${\bf A} \in \mathbb{C}^{I \times F}$ and ${\bf C} \in \mathbb{C}^{P \times F}$ are Vandermonde and that ${\bf B} \in \mathbb{C}^{J \times F}$ is tall and full rank. If,
\begin{align}\label{IdentCondition}
k_{{\bf A}} + \min(P-1,F) \geq F +1. 
\end{align}
then the PARAFAC decomposition of $\mathcal{X}$ in terms of ${\bf A}$, ${\bf B}$, and ${\bf C}$ is essentially unique, where $k_{\bf A}$ denotes the Kruskal rank (k-rank) of the matrix ${\bf A}$.
\end{theorem}
\noindent  It has been shown in \cite{sidiropoulos2000parallel} that a matrix with Vandermonde structure has  full k-rank, i.e., $k_{\bf A} = \min(I,F)$. The condition in \eqref{IdentCondition}  will be interpreted later in the context of hybrid beamforming.

\vspace{-0.15cm}
\section{Hybrid Beamforming via PARAFAC}\label{SubSec:PropMethod}
In this section, it will be shown how the wideband hybrid beamforming problem in ~\eqref{OP:WBHBFVan} can be reformulated as a tensor decomposition problem. Let us define the matrices ${\bf X} = [{\bf F}_\text{opt}[1],\cdots,{\bf F}_\text{opt}[K]] \in \mathbb{C}^{N_t \times KN_s}$ and ${\bf B} = [{\bf F}^T_\text{BB}[1],\cdots,{\bf F}^T_\text{BB}[K]]^T \in \mathbb{C}^{KN_s \times N_t^\text{RF} }$, then it can be easily seen that \eqref{OP:WBHBFVan} can be expressed in more compact form as 
\begin{subequations}\label{OP:WBHBFVanCompact}
	\begin{align}
    &\text{Find} \quad  {\bf F}_{\text{RF}}(\phi_0,\cdots,\phi_{N_t^\text{RF} - 1}),~ {\bf B}  \\
    &\textrm{s.t.} \quad~~ {\bf X} \approx {\bf F}_{\text{RF}}{\bf B}^T, ~{\bf F}_\text{RF} \in \mathcal{F}_V.
	\end{align}
\end{subequations}
\begin{remark}
    Notice that while \eqref{OP:WBHBFVanCompact} assumes a uniform linear array (ULA) structure on the columns of the analog beamformer ${\bf F}_\text{RF}$, the proposed tensor method can be further extended to handle other array structures, for e.g., uniform planar array (UPA) \cite{sidiropoulos2000parallel}. In that sense, the proposed method can be used to recover  azimuth and elevation estimates for each column of ${\bf F}_\text{RF}$. This is in fact a big advantage of the proposed approach relative to the state-of-the-art. Owing to space limitations, we will present only the ULA structure here.
\end{remark}
Let us construct the following two subarrays,  
\begin{subequations}
	\begin{align}
	        &{\bf A} = {\bf F}_\text{RF}(1:end-1,:),~ (\text{all rows except last}) \\
	        &\overline{\bf A} = {\bf F}_\text{RF}(2:end,:),~ (\text{all rows except first})
	\end{align}
\end{subequations}
Then, it follows that by exploiting the Vandermonde structure of the columns of the matrix ${\bf F}_\text{RF}$, the $ {(N_t-1) \times N_t^\text{RF}}$ matrices ${\bf A} $ and $\overline{\bf A}$  are displaced but otherwise identical subarrays, i.e.,
\begin{align}
    \overline{\bf A} = {\bf A}{\bf \Phi}_1.
\end{align}
where ${\bf \Phi}_1 := \text{Diag}([e^{-\phi_0},\cdots,e^{-\phi_{N_t^\text{RF}-1}}])$. Further, for consistency, let ${\bf A} = {\bf A}{\bf \Phi}_0$, where ${\bf \Phi}_0 = {\bf I}_{N_t^\text{RF}}$. Let ${\bf C} \in \mathbb{C}^{P \times N_t^\text{RF}}$ be a matrix holding the diagonal of ${\bf \Phi}_p$ on its $p$-th row, for $p =0,\cdots,P-1$ and $P = 2$. Then, upon defining
${\bf X}_0 = {\bf X}(1:end-1,:) \in \mathbb{C}^{(N_t -1)\times KN_s}$, ${\bf X}_1 = {\bf X}(2:end,:) \in \mathbb{C}^{(N_t -1)\times KN_s}$ and ${\bf D}_p({\bf \Phi}) = {\bf \Phi}_p$, for $p =0,1$, we can write the following,
\begin{align}\label{HBFSlab1}
	        & {\bf X}_0 ={\bf A}{\bf D}_0({\bf C}){\bf B}^T, \\
            & {\bf X}_1 ={\bf A}{\bf D}_1({\bf C}){\bf B}^T \label{HBFSlab2}. 
\end{align}
From the PARAFAC decomposition slab format defined in \eqref{TensorDecomSlab}, it is easy to see that \eqref{HBFSlab1} and \eqref{HBFSlab2} form a two-slab, i.e., $P=2$, PARAFAC model with Vandermonde structure in one mode. Thus, solving \eqref{OP:WBHBFVanCompact} is tantamount to decomposing the tensor $\mathcal{X} \in \mathbb{C}^{(N_t-1) \times KN_s \times P}$ with its $p$-th slab defined as $\mathcal{X}(:,:,p):={\bf X}_p$, for $p=0,\cdots,P-1$ and $P=2$. From an optimization perspective, this can be expressed as
\begin{align}\label{Op:PARAFACDec}
    \min_{{\bf A},{\bf B},{\bf C}}~~ \|\mathcal{X} - \left\llbracket{\bf A},{\bf B},{\bf C}\right\rrbracket\|_F^2
\end{align}
Several algorithms have been developed to tackle the optimization problem \eqref{Op:PARAFACDec} \cite{sidiropoulos2017tensor}. In this work, we adopt the trilinear alternating least square (TALS) algorithm implemented in the widely known Tensorlab MATLAB toolbox \cite{Verv16}.

Considering the condition in \eqref{IdentCondition} in the context of hybrid beamforming, one can easily see that with $P = 2$ and given that ${\bf A}$ is tall (i.e., $N_t \geq N_t^\text{RF} + 1 $) and Vandermonde,  the condition in \eqref{IdentCondition} is always satisfied.
The only requirement though to ensure essential uniqueness of ${\mathcal{X}}$ is that ${\bf B} \in \mathbb{C}^{KN_s \times N_t^\text{RF}}$ needs to be tall and full rank. This requires the number of subcarriers multiplied by the number of streams be greater than or equal to the number of transmit RF chains. This renders our proposed method not applicable for single carrier systems, i.e., $K=1$, with $ N_t^\text{RF} > N_s$,  otherwise, such a condition can be easily satisfied with a modest number of subcarriers.

Let $\overline{\bf A} \in \mathbb{C}^{(N_t-1) \times N_t^\text{RF}}$, $\overline{\bf B}  \in \mathbb{C}^{KN_s \times N_t^\text{RF}}$ and $\overline{\bf C}  \in \mathbb{C}^{2 \times N_t^\text{RF}}$ be the resulting solution of \eqref{Op:PARAFACDec}.  The goal now is to find $(\phi_0,\cdots,\phi_{N_t^\text{RF} - 1})$ and $\{{\bf F}_\text{BB}[k]\}_{k=1}^{K}$ given $(\overline{\bf A},\overline{\bf B},\overline{\bf C})$. To do so, we first recover the $N_t^\text{RF}$ phases $\{{\phi_i}\}_{i=0}^{N_t^\text{RF}-1}$ from the columns of $\overline{\bf A}$ by simply reading the angles of first elements of the columns of $\overline{\bf A}$.

\begin{algorithm} 
	\SetAlgoLined
	\textbf{Input}:  $\{{\bf F}_\text{opt}[k] \in \mathbb{C}^{N_t \times N_s}\}_{k=1}^{K},~$ $N_t^\text{RF}$,\\
	Construct ${\bf X} = [{\bf F}_\text{opt}[1],\cdots,{\bf F}_\text{opt}[K]] \in \mathbb{C}^{N_t \times KN_s}$ \\
	Construct $\mathcal{X} \in \mathbb{C}^{(N_t - 1) \times KN_s \times 2}$ as $\mathcal{X}(:,:,1) = {\bf X}(1:end-1,:)$, and  $\mathcal{X}(:,:,2) = {\bf X}(2:end,:)$ \\
	Decompose $\mathcal{X} := \left\llbracket \overline{\bf A},\overline{\bf B},\overline{\bf C}\right\rrbracket$ using TALS \\
	\For{$ i = 0:N_t^\text{RF}-1$}{
	 Recover $\phi_i$ by computing the angle of the first element of $\overline{\bf A}(:,i)$ \\
	 Form ${\bf F}_\text{RF}(:,i) = [1,e^{j\phi_i},\cdots,e^{j(N_t -1)\phi_i}]^T$ \\
	 Obtain $\lambda^{i}_1 = \overline{\bf A}(1,i)$, $\lambda^{i}_3 = \overline{\bf C}(1,i)$, $\lambda^{i}_2 = \frac{1}{\lambda^{i}_1\lambda^{i}_3}$ \\ 
	 Obtain ${\bf B}(:,i) = \overline{\bf B}(:,i)/\lambda^{i}_2$}
	 Reshape ${\bf B}$ to retrieve $\{{\bf F}_\text{BB}[k] \in \mathbb{C}^{N_t^\text{RF} \times N_s}\}_{k=1}^K$.
	\caption{V-TPAR: Vandermonde Two-slab PARAFAC}
	\label{Algo1}
\end{algorithm}  
To obtain $\{{\bf F}_\text{BB}[k]\}_{k=1}^{K}$, we need to resolve the complex scaling ambiguity that is inherent to PARAFAC (see Definition 1 for the essential uniqueness of PARAFAC). Note that we ignore the permutation ambiguity, as in the hybrid beamforming context, finding the analog and baseband precoders up to a common permutation ambiguity is irrelevant since it merely amounts to shuffling the RF chains. The complex scale ambiguity though is important as it amounts to entirely changing the directions of the precoders. Fortunately, since the columns of both matrices $\overline{\bf A}$ and $\overline{\bf C}$ exhibit a Vandermonde structure, the column-wise scale ambiguity in both matrices can be resolved by simply dividing the elements of each column by the first element. Once the complex scale ambiguities associated with the columns of  $\overline{\bf A}$ and $\overline{\bf C}$, denoted as ${\bf \Lambda}_1$ and ${\bf \Lambda}_3$, respectively, are resolved, it can be seen from \eqref{TensPermutaionandScaling} that the column-wise scale ambiguity of ${\bf B}$, denoted as ${\bf \Lambda}_2$, can be easily obtained as ${\bf \Lambda}_2 = ({\bf \Lambda}_3{\bf \Lambda}_1)^{-1}$. The above procedures for solving the wideband hybrid beamforming problem using Vandermonde-constrained Two-slab PARAFAC (V-TPAR) are outlined in Algorithm \ref{Algo1}.

The complexity of Algorithm 1 is incurred in decomposing the tensor $\mathcal{X}$ using the iterative TALS algorithm. The per iteration complexity of TALS is equal to the cost of inverting an $(N_t^\text{RF})^2 \times (N_t^\text{RF})^2 $ matrix. The overall complexity then depends on the total number of iterations which in turn depends on the problem and the size of the tensor (see \cite{sidiropoulos2017tensor} and references therein for convergence properties of TALS).  As we will see later, for the considered problem, a few iterations of TALS seem to be sufficient to obtain hiqh-quality solution.

  \begin{figure}
 	\centering
 	\includegraphics[width=2.6in]{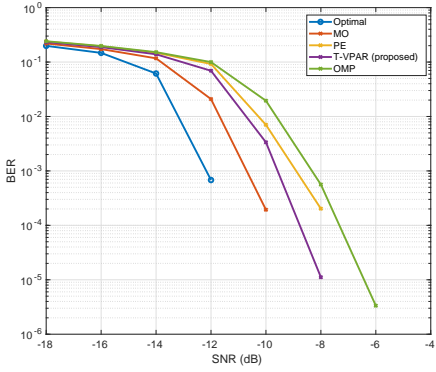}
 	\caption{The BER  performance with $ N^\text{RF} = 2 $ and $N_s = 1$.}
 	\label{Fig:BER2x1}
 \end{figure}

 \begin{table}
\centering
    \begin{tabular}{|p{3.5cm}|p{2cm}|}
        \hline
       ~~~~ Parameter                 & ~~~~Value\\
        \hline
      Carrier frequency       &   28 GHz  \\
      Subcarrier spacing       & 60 kHz \\
      Modulation             & 16-QAM \\
      Code rate              & 0.49 \\
      Number of transmit antennas    & 32 \\
      Number of receive antennas     & 8 \\
      UE speed             & 0.5 km/hr \\
      Delay spread             & 300 ns \\
      Channel model              & CDL-C \\
      \hline
    \end{tabular}
        \caption{Parameter settings for the simulations.}
    \label{Table:Simu_parameters}
\end{table}
\vspace{-0.3cm}
\section{Simulations}\label{Sec:Simu}
In this section, we will provide numerical results on 3GPP link-level channel model to assess the performance of the proposed method. The adopted simulation parameters are listed in Table \ref{Table:Simu_parameters}. We use the CDL-C channel model with the delay spread set to 300 ns. Both BS and UE are equipped with uniform linear array where the antenna elements are separated by a half wavelength. All results are averaged out over 200 realizations. The number of subbands is set to $30$, i.e., $K = 30$, where each subband consists of one resource block (RB), i.e., 12 subcarriers. The channel matrix for each subband is obtained by averaging out the channels across the 12 subcarriers. For the proposed method implementation, we used the TALS algorithm implemented in the Tensorlab MATLAB toolbox. Finally, all simulations were performed on an Intel(R) Xeon(R) Gold 6234 CPU.

\begin{figure}
 	\centering
 	\includegraphics[width=2.8in]{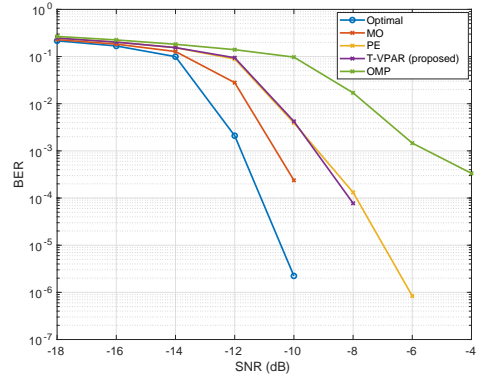}
 	\caption{The BER performance $ N^\text{RF} = 3 $ and $N_s = 2$.}
 	\label{Fig:BER3x2}
 \end{figure}

To benchmark the performance of the proposed method, we use the manifold optimization (MO) alternating minimization algorithm \cite{yu2016alternating}, the phase extraction (PE) alternating minimization algorithm \cite{yu2016alternating} and the OMP algorithm \cite{el2014spatially} as baselines. Both MO and PE solve the wideband hybrid beamforming problem \eqref{OP:WBHBF} with unit modulus constraints on the entries of the analog beamformers, while the OMP algorithm solve \eqref{OP:WBHBF} with codebook constraint on the columns of the analog beamformers. For OMP, we use the DFT codebook for both ${\bf F}_\text{RF}$ and ${\bf W}_\text{RF}$.

From the feedback overhead perspective, one can see from Table \ref{Table:FeedbackOverhead} that  the Vandermonde feasible set (our proposed method) attains the same overhead of the codebook one (OMP). In particular, the number of parameters to feed back is independent of the number of transmit (receive) antennas and is equal to the number of transmit (receive) RF chains if the the analog precoders and combiners are computed at the UE (BS). On the other side, the unit-modulus feasible set (MO and PE) suffers from the large overhead that scales up with the number of transmit/receive antennas, thereby limiting their use in limited feedback systems.  
\begin{table}
\centering
    \begin{tabular}{|p{2.3cm}|p{1.6cm}|p{1.4cm}|p{1.4cm}|}
        \hline
      ${\bf F}_\text{RF}$ feasible set &  ~~~  $\mathcal{F}_u$ & ~~~ $\mathcal{F}_c$& ~~~$\mathcal{F}_v$\\
      \hline
     Num. of parameters    & ~~~~$N_t N_t^\text{RF}$ & ~~~~$N_t^\text{RF}$ & ~~~~$N_t^\text{RF}$ \\
      \hline
     Method    & ~~~~MO and PE & ~~~~ OMP & ~~~~T-VPAR \\
     \hline
    \end{tabular}
        \caption{Feedback overhead associated with the different feasible sets of the analog beamformers.}
    \label{Table:FeedbackOverhead}
\end{table}

To evaluate the practical impact of the different hybrid beamforming algorithms, we report the coded BER in an end-to-end system. First, we consider a scenario with $N^\text{RF} = 2$ and $N_s = 1$, i.e., $N^\text{RF} = 2N_s$, while the rest of the parameters are as listed in Table \ref{Table:Simu_parameters}. It is known from \cite{yu2016alternating} that when $N^\text{RF}=N_s$, PE achieves the same performance of MO at much lower complexity while the performance of the former degrades when $N^\text{RF} > N_s$.  Fig. \ref{Fig:BER2x1} shows the end-to-end coded BER performance of the different methods. One can see that, for this case, the proposed method achieves more than 1 dB SNR gain relative to OMP. More interestingly, the proposed approach outperforms the PE method with more than an order of magnitude reduction in BER at -8 dB.  Further, when $N_s < N^\text{RF} < 2N_s$ as shown in Fig. \ref{Fig:BER3x2}, the performance of the proposed method significantly outperforms OMP with roughly 4 dB SNR gain. Finally, one can see that both the tensor method and PE attain approximately the same performance, with 1 dB loss relative to MO.

\begin{figure}
 	\centering
 	\includegraphics[width=2.6in]{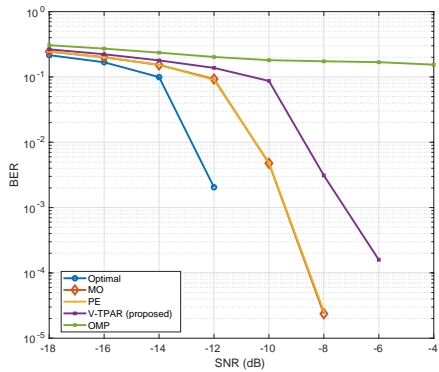}
 	\caption{The BER performance $N^\text{RF} = N_s = 2$.}
 	\label{Fig:BER2x2}
\end{figure}

Next, we simulated another scenario with $N^\text{RF} = N_s = 2$. It can be seen that now PE achieves the same performance as MO while the proposed method incurs roughly 2 dB SNR loss, as Fig. \ref{Fig:BER2x2} depicts. In addition, one can see that the proposed algorithm considerably outperforms the OMP algorithm with more than an order of magnitude reduction in BER when the SNR exceeds -10 dB.  

\begin{figure}
    \begin{subfigure}[b]{0.23\textwidth}  
        \centering 
        \includegraphics[width=\textwidth]{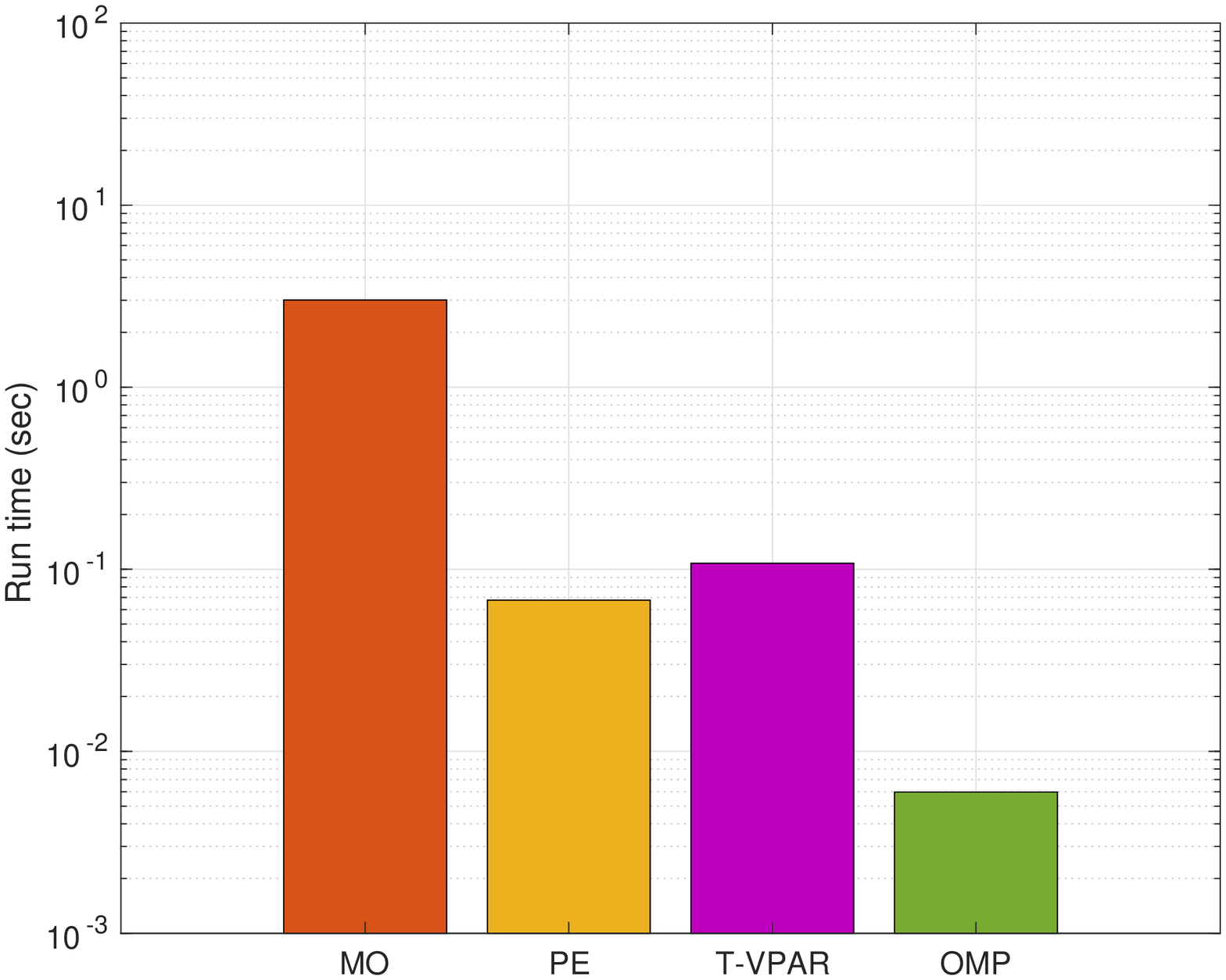}
        \caption{{\small $N^\text{RF} = N_s = 2$ }}    
        \label{Fig:time2x2}
    \end{subfigure}
    \hspace{0.1cm}
    \begin{subfigure}[b]{0.23\textwidth}   
        \centering 
        \includegraphics[width=\textwidth]{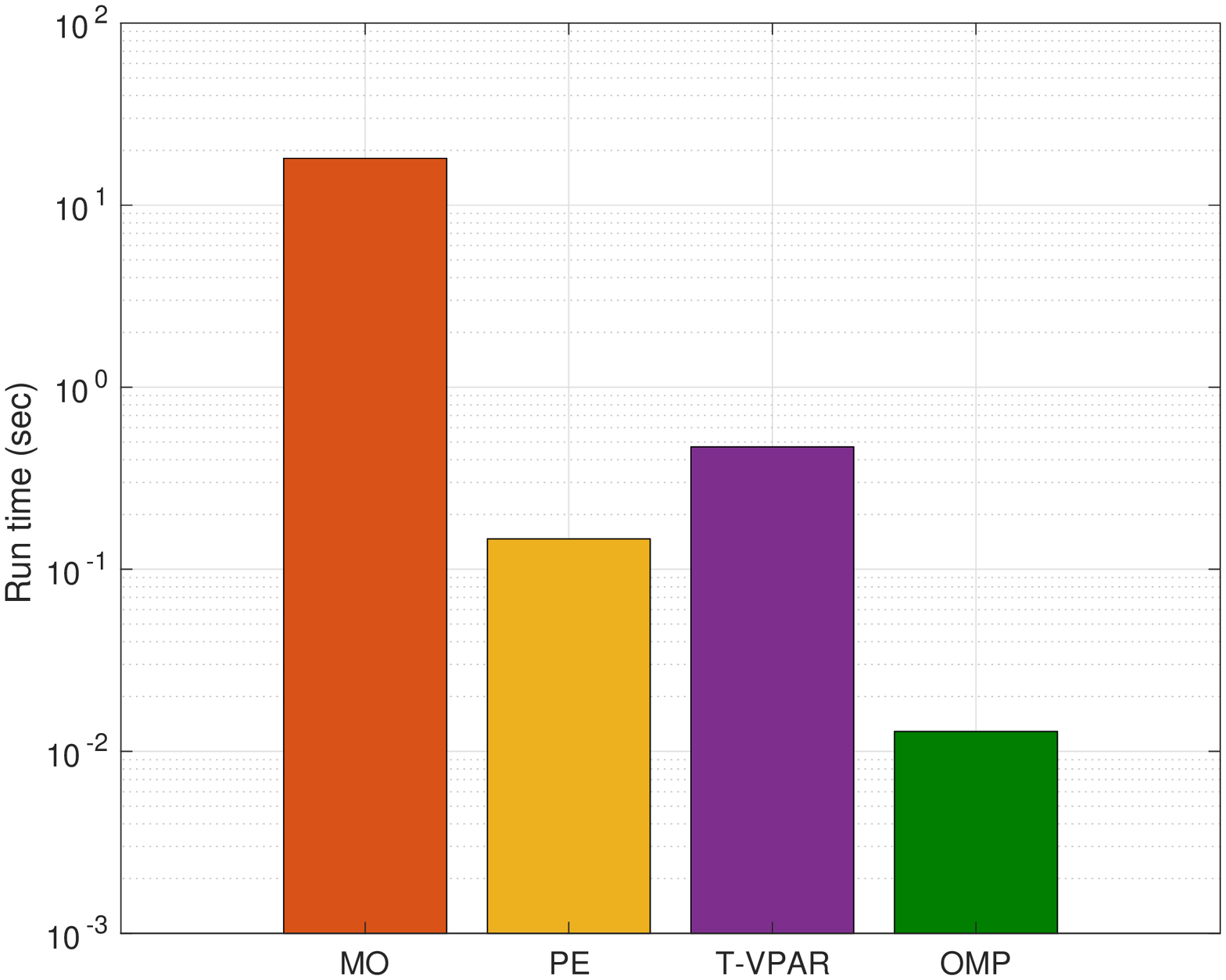}
        \caption {{\small $N^\text{RF} = 3$ and $N_s = 2$}}    
        \label{Fig:time3x2}
    \end{subfigure}
    \caption{\small Average run time results.} 
    \label{fig:time}
\end{figure}

Finally, to assess the complexity of the proposed tensor approach, Fig. \ref{fig:time} depicts the average run time of the proposed method relative to the considered baselines, when $N^\text{RF} = N_s = 2$ and $N^\text{RF} = 3$ and  $N_s = 2$ . We observe that the run time of the proposed method is comparable to PE while achieving more than an order of magnitude reduction in run time compared to MO in both setups. Finally, OMP features the lowest run time but this obviously comes at the expense of performance.

\vspace{-0.5cm}
\section{Conclusions}\label{Sec:Conc}
This paper has considered single user hybrid precoding and combining in wideband mmWave MIMO systems under Vandermonde constraints on the hybrid precoders and combiners. The problem is formulated as a tensor factorization problem where PARAFAC is invoked to find the Vandermonde-constrained analog beamformers and the set of baseband precoders -- with identifiability guarantees. Numerical results on a 3GPP link-level test bench have revealed the superiority of the proposed method relative to the state-of-the-art. In particular, the proposed method has shown to be striking the balance between performance, overhead and complexity. As a future work, we aim at expanding the applicability of the proposed framework to other array structures such as uniform plannar array (UPA). Further, we plan to explore the impact of increasing the number of subarrays (multi-slab PARAFAC as opposed to two slab) on the estimation accuracy, and its trade-off with  computational complexity.

\bibliographystyle{IEEEtran}
\bibliography{IEEEabrv,references}

\end{document}